*Manuscript

# Towards a better understanding of the structure of diamanoïds and diamanoïd/graphene hybrids

Fabrice Piazza[1*], Marc Monthioux[2], Pascal Puech[2], Iann C. Gerber[3]

[1] Nanoscience Research Laboratory, Pontificia Universidad Católica Madre y Maestra, Autopista Duarte km 1 1/2, Apartado Postal 822, Santiago, Dominican Republic

[2] Centre d'Elaboration des Matériaux et d'Etudes Structurales (CEMES), CNRS, Université de Toulouse, 29, rue Jeanne Marvig, BP 94347, 31055 Toulouse Cedex 4, France

[3.] Laboratoire de Physico-Chimie des Nano-Objets (LPCNO), CNRS, INSA, Université de Toulouse, 135 Avenue de Rangueil, 31400 Toulouse, France

## Abstract

Hot-filament process was recently employed to convert, totally or partially, few-layer graphene (FLG) with Bernal stacking into crystalline $sp^3$–C sheets at low pressure. Those materials constitute new synthetic carbon nanoforms. The result reported earlier relies on Raman spectroscopy and Fourier transform infrared microscopy. As soon as the number of graphene layers in the starting FLG is higher than 2-3, the $sp^2$–C to $sp^3$–C conversion tends to be partial only. We hereby report new evidences confirming the $sp^2$–C to $sp^3$–C conversion from electron diffraction at low energy, Raman spectroscopy and Density Functional Theory (DFT) calculations. Partial $sp^2$–C to $sp^3$–C conversion generates couples of twisted, superimposed coherent domains (TCD), supposedly because of stress relaxation, which are evidenced by electron diffraction and Raman spectroscopy. TCDs come with the occurrence of a twisted bilayer graphene feature located at the interface between the upper diamanoïd domain and the non-converted graphenic domain underneath, as evidenced by a specific Raman signature consistent with the literature. DFT calculations show that the up-to-now poorly understood Raman T peak originates from a $sp^2$–C-$sp^3$–C mixt layer located between a highly hydrogenated $sp^3$–C surface layer and an underneath graphene layer.

*Corresponding author. E-mail: fpiazza@pucmm.edu.do (Fabrice Piazza)

## 1. Introduction

Since 2009, diamane has been a hypothetical nanocarbon material [1]. Interest in genuine diamane (two $sp^3$–C layers only, both hydrogenated) and more generally in diamanoïds [2], which are wide band-gap 2D materials, derives from its potential use in technologies such as transistors for nanoelectronics, host system for single photon emission in quantum computing devices and building material for miniaturized biomedical devices [1,2,3,4,5,6]. Currently, the ability to create graphene transistors is limited by the lack of band-gap in that material [7]. Unlike graphene, according to computational studies, diamanoïds display the electronic structure of semiconductors, with a direct band-gap depending on the film thickness and crystalline structure (stacking order) [1,3,4]. Another interest lies in the possibility of using nitrogen-vacancy (N–V) centers in diamane through substitution by nitrogen atoms in the C lattice to configure qubits for use in quantum computing [5,6]. Diamane was recently expected to be superior to diamond as host system for single photon emission [6]. Finally, the expected strength, low coefficient of friction and bio-compatibility could render diamane and diamanoïds very competitive as building materials to make lower power and miniaturized electronics and biomedical devices [2].

In 2019, the original synthesis of few-layer and crystalline $sp^3$–carbon (lonsdaleite or diamond or a combination of both) was demonstrated over surface areas up to ~1,700 $\mu m^2$ [2]. This breakthrough was achieved from the chemisorption of H radicals produced by the hot-filament process at low temperature (below 325 °C) and pressure (50 Torr) on the basal planes of few-layer graphene and the subsequent interlayer bonding between $sp^3$–C. Until then, even if various studies based on calculations hypothesized that diamanoïds could form when non-supported few-layer graphene (FLG) are placed in hydrogen cold plasma [1], there was no report of any experimental evidence of such scenario. A structure transformation of FLG into a diamond-like layer was reported in [8]. However, crystal structure was not evidenced. No Raman spectroscopy nor electron diffraction results were reported. The material obtained in [8] could be $sp^3$–C-rich amorphous carbon, that is to say tetrahedral amorphous carbon [9,10]. Temporary interlayer bonding between graphene layers were reported but this was shown to happen only when graphene layers are exposed to very high pressure [11,12,13]. In contrast, in [2], it was unambiguously shown, for the first time, the preparation at low pressure of stable nanometer-thick crystalline $sp^3$–bonded carbon materials. $sp^3$–C related peaks from diamond and/or lonsdaleite and/or hybrids of both were detected in visible and UV Raman spectra. C-H bonding was directly detected by Fourier Transform Infrared (FTIR) microscopy over an area of ~150 μm$^2$ and one single component attributed to *sp*$^3$–C–H mode was detected in the C–H stretching band showing that carbon was bonded to one single hydrogen and strongly suggesting that the *sp*$^3$–C materials obtained were nanometer-thick films with basal planes hydrogenated [2]. However, the *sp*$^2$–C to *sp*$^3$–C conversion was not confirmed by transmission electron microscopy (TEM) techniques, because the converted material was found to be electron sensitive, even for electron energy as low as 80 keV, supposedly because of the high hydrogenation degree of the surface, which was evidenced by FTIR microscopy [2]. Following the publication of [2], another possibility for the pressureless *sp*$^2$–C to *sp*$^3$–C transformation of nanometer-thick and crystalline films, from fluorine chemisorption, was submitted for publication [14]. In [14], the disappearance of *sp*$^2$–C features in Raman spectra following fluorine chemisorption was shown although sp$^3$–C features were not evidenced in Raman spectra. However, some evidences of the *sp*$^2$–C-to-*sp*$^3$–C conversion were provided by optical transmission and transmission electron microscopy techniques. The most convincing evidence is the disappearance of the 1s–2p ($\pi^*$) energy loss peak at 285 eV in the high energy loss spectra of fluorinated bi-layer graphene obtained at 80 keV. The electron sensitivity of the film was shown at 80 kV by varying the size of the parallel electron beam and hence the electron dose rate. Notably, there is a lack of overlap between the experimental results published in the so far only two reports on nanometer-thick and crystalline *sp*$^3$–C sheets [2,14]. Further structure analysis is required.

In this work, we studied the structure of nanometer-thick and crystalline *sp*$^3$–C sheets obtained by the exposure of FLG to H radicals in a hot-filament chemical vapor deposition (HFCVD) reactor, before and after hydrogenation, by TEM and electron diffraction at 5 keV, Density Functional Theory (DFT) calculations, and further interpretation of Raman spectra.

## 2. Experimental

### 2.1. Graphene pristine materials

As-received FLG films deposited on 3 mm copper transmission electron microscopy grids (2000 Mesh) from Graphene Supermarket (SKU # SKU-TEM-CU-2000-025) were used as graphene materials. The FLG films were grown by CVD from $CH_4$ at 1000°C on Ni substrate and transferred onto a commercial TEM grid using a polymer-free transfer method to minimize contamination as described in Ref. [15]. FLG thickness is typically between 0.3-2 nm (1-6 monolayers), however thicker film, up to 20 layers, were observed in TEM. Typical FLG TEM grid coverage is between 60 and 90 %.

### 2.2. Hydrogenation process

A commercial hot filament reactor previously described in details [16], was used for hydrogenation and subsequent structure conversion of FLG into nanometer-thick and crystalline $sp^3$–C sheets. The details of the process were previously disclosed [2]. The samples studied here were prepared at a pressure of 10 Torr and using a $H_2$ flux of 1 sccm. The duration of the hydrogenation process was of 9 minutes and 50 seconds. The maximum FLG temperature was ~325 °C [2].

### 2.3. Material characterization

The $sp^2$–C to $sp^3$–C conversion was confirmed by UV Raman spectroscopy following the work published in [2]. For that purpose, UV Raman spectroscopy was employed to examine the material structure before and after hydrogenation. Raman spectra were recorded with a Renishaw InVia Reflex Spectrometer System equipped with a stigmatic single pass spectrograph including holographic grating of 3600 grooves.mm$^{-1}$ and using the 244 nm lines of Ar ion laser (85 Lexel Second Harmonic Generation laser). The scattered light was collected in the 180° backscattering geometry. A x40 objective was employed. The detection system was a Peltier-cooled UV coated Deep Depletion CCD array detector and the entrance slit was set to 50 μm. The resolution was of 3.5 cm$^{-1}$. The laser power on the sample and acquisition time were adjusted to obtain optimum signal without any sample modification. Typically, laser power was in the range of 1 mW, and exposure time to the laser was in the range of 1 s. No visible damage and no change of the spectral shape during measurements have been observed. Highly-oriented pyrolytic graphite (HOPG) was used for peak position calibration. Raman mapping was employed using high-speed encoded mapping stage and a 1” CCD to generate high definition 2D chemical images over thousands of square microns.

Probed surface area was typically of ~140×100 μm$^2$. Peaks were fitted using Lorentzian function using Wire 4.4 software from Renishaw.

TEM and electron diffraction were employed to further examine the material structure before and after hydrogenation. It is worth noting that the converted material, as opposed to the behavior of graphene or diamond, appeared to be electron sensitive, possibly because of the high hydrogenation degree of the surface, even for electron energy as low as 80 keV [2]. This is consistent with the photon sensitivity observed under certain conditions [2] and the reported electron sensitivity of F-diamane [14]. This is the reason why a low-voltage benchtop transmission electron microscope at 5keV from America Delong was used. The instrument includes a 5kV Schottky type field emission gun and a 2560×2160 pixel Front Illuminated Scientific CMOS (6.5 μm$^2$ pixel size). Maximum magnification and resolution were of ×700,000 and 1.2 nm, respectively. Electron diffraction patterns were obtained from 100 nm-large areas, and they were analyzed using ImageJ 1.52a software.

Electron diffraction patterns have been calculated using the XaNSoNS open source software considering a wavelength of 0.01nm [17]. Typically $10^6$ atoms have been considered for a bilayer domain and $2x10^6$ atoms for a 4-layer domain. The results have been plotted using Matlab® software.

### 2.4. Computational Details

The electronic structures were obtained from DFT calculations with the VASP package [18,19] which uses the plane-augmented wave scheme [20,21] to treat core electrons. Perdew-Burke-Ernzerhof (PBE) functional [22] was used as an approximation of the exchange-correlation electronic term for all the geometry optimization steps as well as for phonons and frequencies calculations. The cut-off energy was set to 400 eV, with a Gaussian smearing of 0.05 eV width for partial occupancies. During geometry optimization step all the atoms were allowed to relax with a force convergence criterion below 0.005 eV/Å, using the van der Waals corrected scheme of Grimme *et al.* [23]. A vacuum height of 20 Å was used to avoid spurious interaction between periodic images of the different slabs. To estimate phonon dispersion (7x7) supercells were used with a 7x7x1 grid for k-points sampling, in the Density Functional Perturbation Theory (DFPT) framework, using the Phonopy code [24]. To validate the present computational setting, in **Figure**

**SI.1**, the phonon dispersion curve of a pristine graphene monolayer is given, in which the characteristic G band is located at 1572 $cm^{-1}$.

To mimic our experimental FLG samples which are made of a various number of layers, we have modelled partially hydrogenated systems consisting of four distinct carbon layers (**FIG. 1a**).

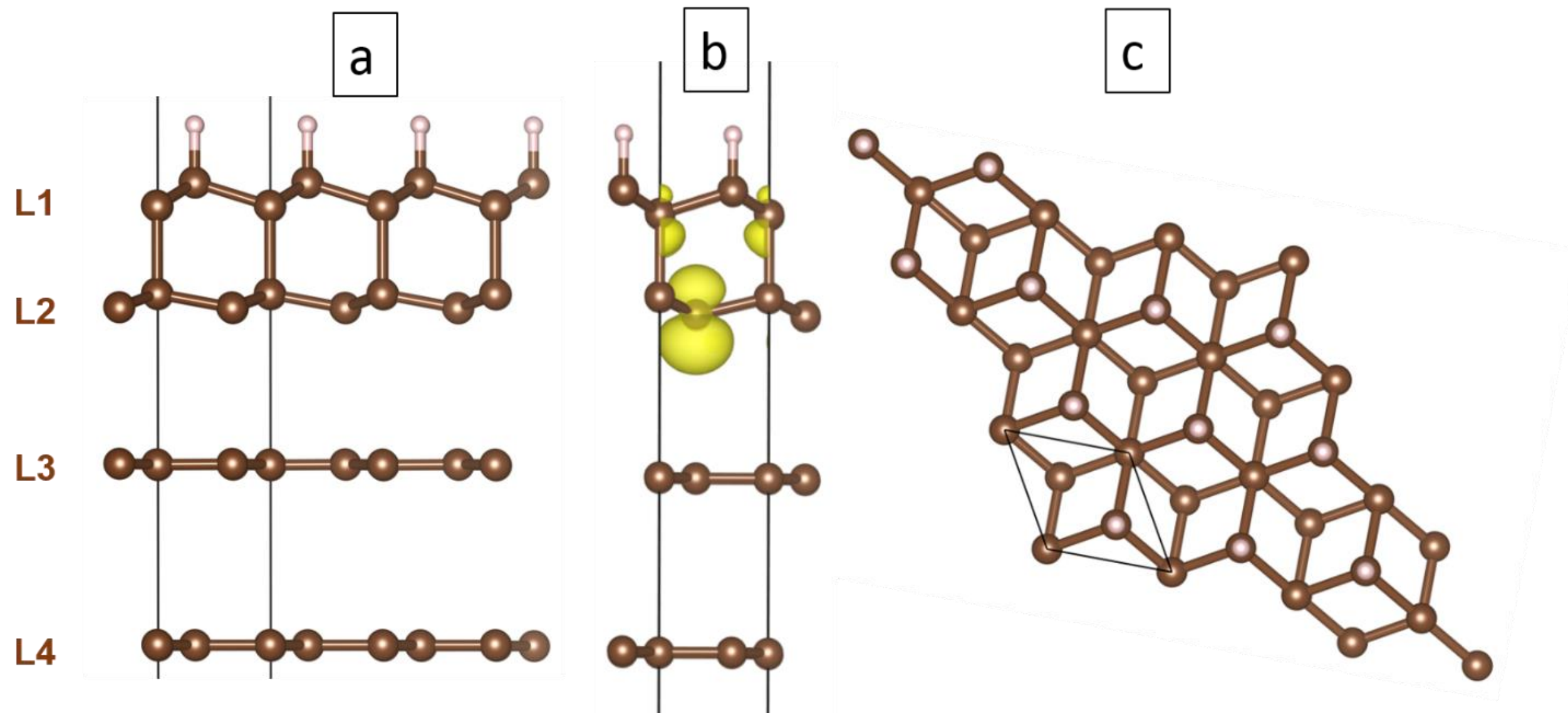

**FIG. 1. (a)** and **(b)** two side view and **(c)** top view (right) of the partially hydrogenated FLG used in DFT calculations with ABBA stacking. The projected structure is that of the face centered cubic of diamond. In **(b)** is shown how the $p_z$ orbitals are preferably oriented towards the layer underneath (L3). The black lines delimite the primitive cell.

The first layer (L1) corresponds to the fully hydrogenated layer typical of diamane and diamanoïds (all C atoms are $sp^3$-hybridized, half of them bonded with one H atom, the other half are bonded with carbon atoms from layer L2). The second layer (L2) is not hydrogenated, hence half of the carbon atoms are truly $sp^3$-C and bonded to carbon atoms from L1, while the other half are still $sp^3$-hybridized but exhibit a free orbital with almost one unpaired electron mainly of $p_z$ character (**FIG. 1b**). The two bottom layers L3 and L4 remain pure $sp^2$–C, however L3 is interacting more strongly with L2 than with L4 due to the existence of the unsatisfied valence bonds in the former. As a consequence, the L2-L3 interlayer distance is shorter (in the range 0.3037-0.3308 nm) than the L3-L4 interlayer distance (0.3512 nm). For thicker systems, other layers similar to L4 can be added underneath while respecting the stacking sequence in graphite (ABAB), until the other side of the flake is reached where the occurrence of the L1-L2-L3 combination may repeat. Despite the stacking sequence of graphene layers in our pristine FLG is likely, it is important to figure out that the $sp^3$–C to $sp^2$–C conversion generates a huge stress in the system, which may induce layer decoupling upon relaxation, more preferably between layers where the interaction is weaker (typically between L3 and L4). Therefore, different L1-…-L4 stacking orders have been tested, namely ABBB, ABAB, ABCA, and ABBA, the latter being more stable by at least 15 meV/carbon atom than the other stacking sequences. As a consequence, only the ABBA stacking sequence is discussed below.

Therefore, the model considered in **FIG. 1** is based on the idea that, when starting from multilayer graphene, the hydrogenation process generates the formation of an upper diamanoïd domain (involving L1 and L2 in the model, possibly added with L3, because of the stronger interaction with L2 than with L4), lying upon a lower, untransformed FLG domain (involving L4 in the model and subsequent graphene layers underneath).

## 3. Results and discussion

**FIG. 2** shows a representative UV Raman spectrum of FLG exposed to H radicals as already reported in [2], which was demonstrated as typical of nanometer-thick and crystalline $sp^3$-C sheets [2]. It is different from the spectra obtained before the hydrogenation process, which are typical of perfect FLG [2]. The spectrum shown in **FIG. 2** displays two sharp peaks centered at ~1062.8 cm$^{-1}$ and 1321.9 cm$^{-1}$, respectively, which the FWHMs are of ~14.0 and 29.6 cm$^{-1}$, respectively. The first peak was interpreted as the *T* peak, somehow related to bonding between $sp^3$-C [2]. This will be further discussed below. The peak centered at ~1321.9 cm$^{-1}$ was interpreted as corresponding to the diamond/lonsdaleite stretching mode [2]. Areas of the specimen for which such peaks could be detected in UV Raman spectra were considered for further TEM examination.

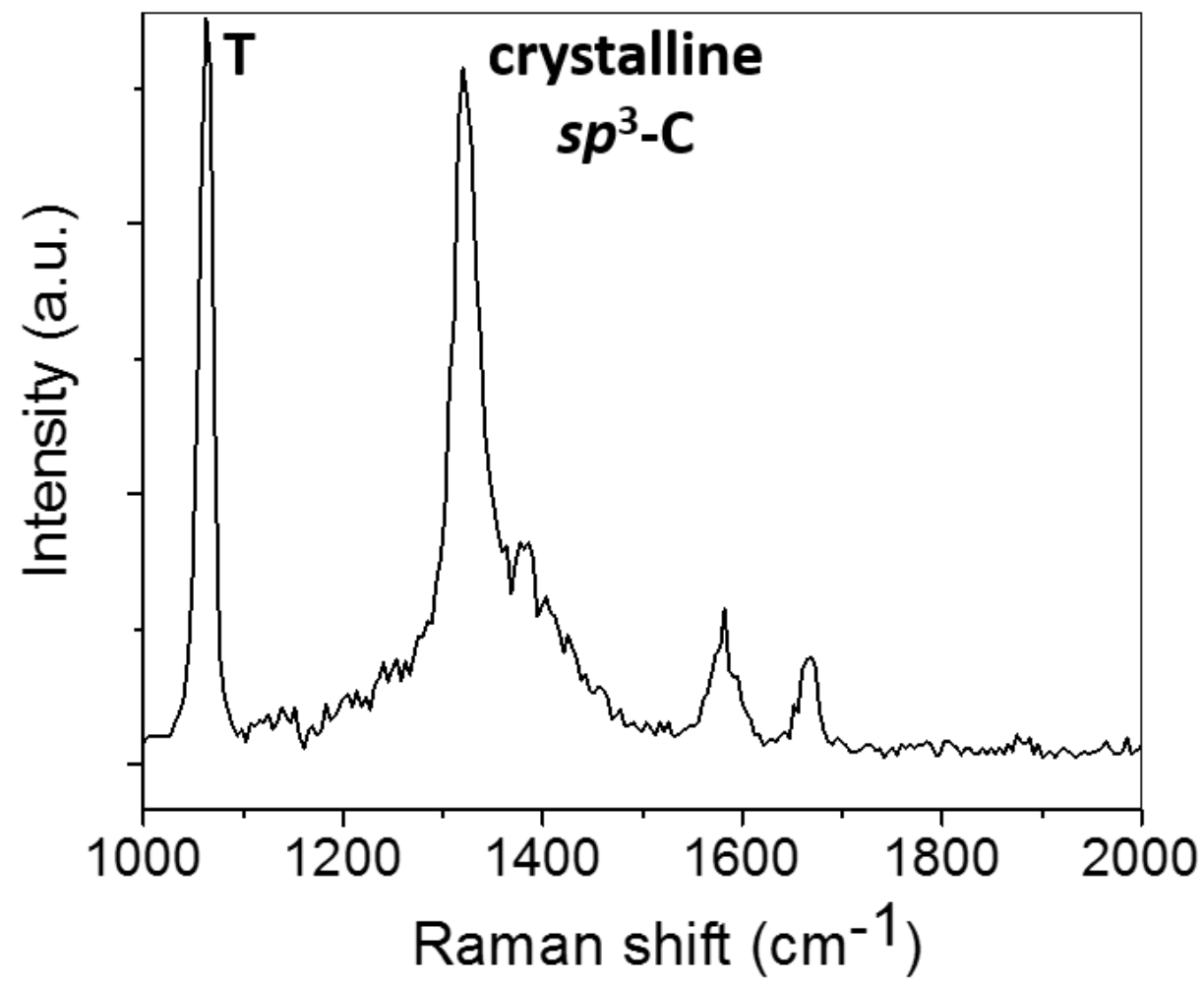

**FIG. 2.** Typical UV Raman spectrum (at 244 nm) of FLG after exposure to the hot-filament-promoted hydrogenation process.

Electron diffraction patterns of FLG before exposure to hydrogenation process were examined. **FIG. 3a** to **3c** show three examples of such patterns taken from three different flakes. **3a** corresponds to a single domain (i.e., with the first and second rings bearing only 6 spots each), while **3b** and **3c** are examples of flakes comprising two and five coherent graphenic domains superimposed, respectively. Over 53 flakes analyzed, only 6 were found to exhibit the diffractogram of a single domain such as in **3a**.

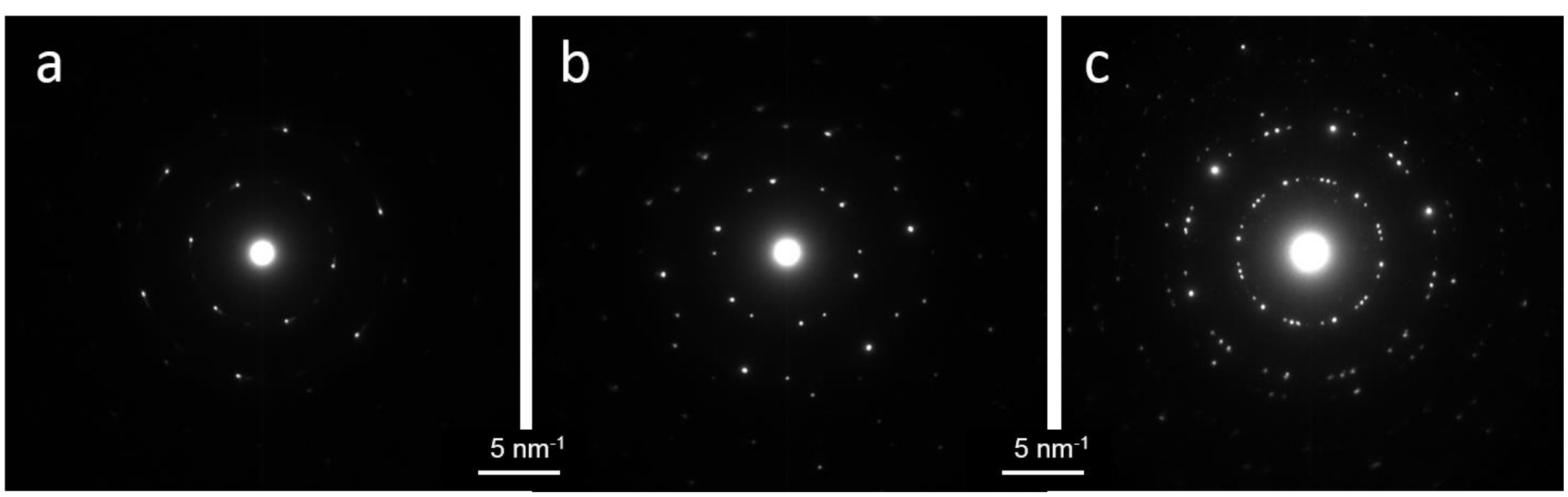

**FIG. 3. (a)** to **(c),** typical electron diffraction patterns of pristine FLG. With respect to the hexagonal structure of graphite, the first ring is the location of the six *100* spots (for a single domain) corresponding to a d-spacing of 0.213 nm, the second ring is the location of the 6 *110* spots corresponding to a d-spacing of 0.123 nm.

In the pristine FLG from the starting specimen (i.e., before hydrogenation), hence multilayer domains, the Bernal stacking sequence ABA is likely, as the most frequent and the most stable (as compared to AAA and ABCA stacking). Hence, the diamond structure (face centered cubic, FCC) should be favored, however, it cannot be ruled out that the huge stresses induced by the $sp^2$–C→$sp^3$–C conversion (see below) may enable atom displacements and generate the lonsdaleite structure (hexagonal) anyway. Therefore, whether it would be possible to discriminate between remaining graphene domain, a converted film with the diamond structure, and the lonsdaleite structure from the diffraction patterns is an important issue to look at. Diffraction pattern may offer three ways for this: interplanar distances, symmetries (spot distribution), and peak intensities.

In order to check whether measuring d-spacing distances might be a reliable discrimination method, the variation in the d-spacing of the (100) planes was obtained from 298 measurements taken from 53 flakes (before hydrogenation). The same magnification (camera length) and the same illumination area were used to record each pattern. In all cases, the distance between opposite spots was considered to minimize measurement errors. The variation in the d-spacing measurement of the (100) planes (theoretical: 0.2130 nm) was found to reach 5.25 %. This value is higher than the difference between the d-spacing of the (111) planes in diamond (0.205 nm [25]) and of the (100) planes in graphite, which is of 3.75 %. Likewise, the d-spacing of the (110) planes in diamond (0.126 nm [25]) is very close of that of the (110) planes in graphite (0.123 nm), which makes a difference of ~2.4%. Similarly, the difference between the d-spacing of the (100) planes in lonsdaleite (0.219 nm [25]) and in graphite is only 2.8 %. Therefore, under the present experimental conditions, it is not possible to unambiguously discriminate the diffraction patterns of pure $sp^3$–C crystalline phase (diamond or diamane or lonsdaleite) from that of few-layer graphene by estimating d-spacing from the diffraction spots.

On the other hand, considering the spot distribution on the diffraction patterns should, in principle, be able to discriminate between the FCC structure of diamond on the one hand, and the hexagonal structures of lonsdaleite and graphene on the other hand. Indeed, as the diamond structure in a converted film is seen according to the [111] axis (perpendicular to the film plane, see **FIG. 1c**), there should be no spot on the first ring, and the first family of spots showing the hexagonal symmetry should be that of the (110) planes, i.e. the 220 spots, corresponding to a d-spacing of 0.126 nm, which locate on the second ring. This is what is shown on the simulated diffraction pattern of a diamond nanoparticle in **FIG. 4a**.

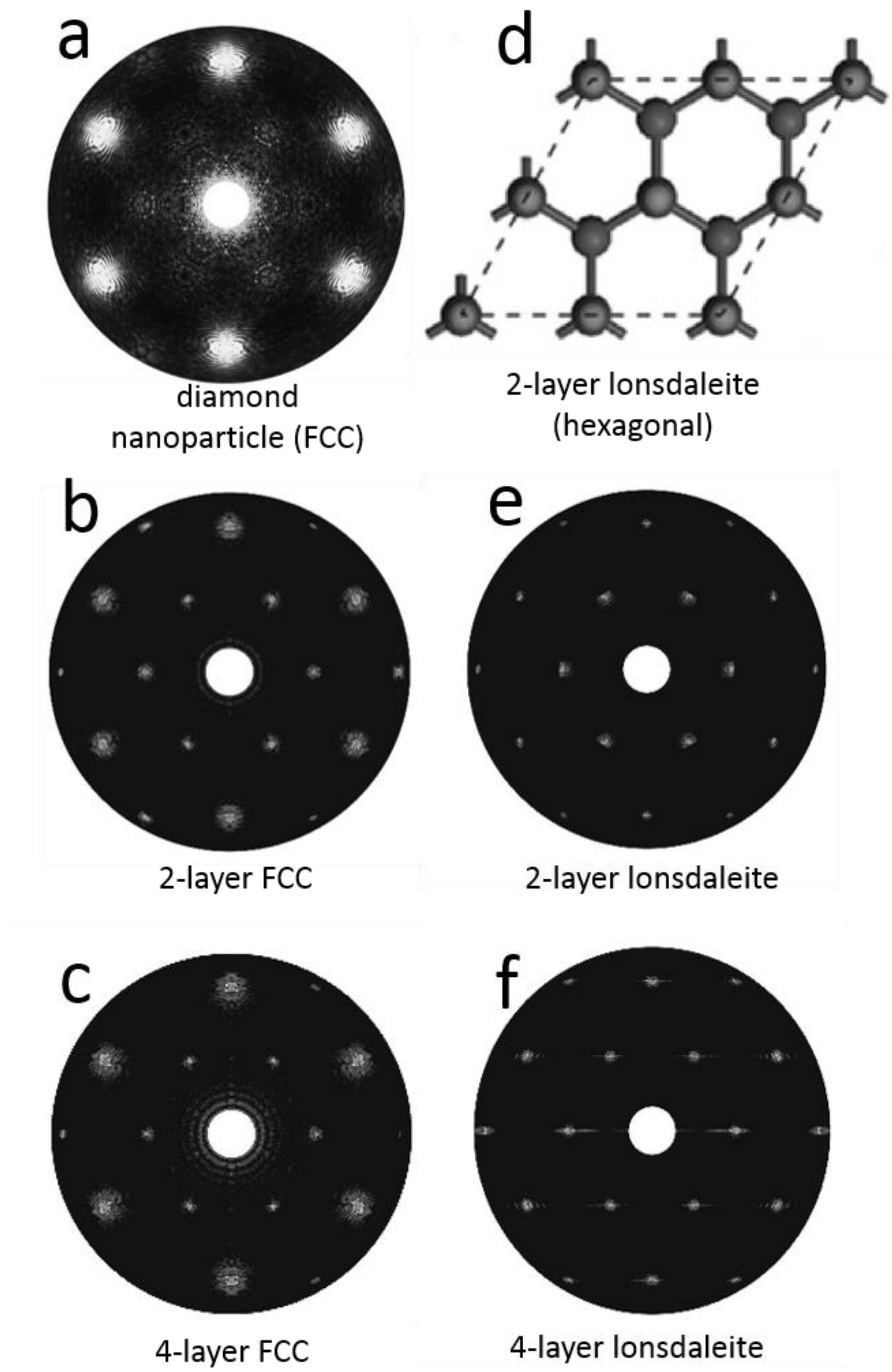

**FIG. 4. (a)** to **(c),** simulated electron diffraction patterns for the FCC diamond structure oriented with the [111] axis parallel to the electron beam. **(a)** Diamond nanoparticle (CFC structure) ; **(b)** 2-layer film ; **(c)** 4-layer film. **(d)** Model of a 2-layer hexagonal structure oriented with the [001] axis parallel to the electron beam ; **(e)** and **(f)** simulated electron diffraction patterns for a 2-layer and a 4-layer film with the hexagonal structure oriented as in **(d)**, respectively. For **(a)**, **(b)**, and **(e)** $10^6$ atoms were considered, for **(c)** and **(f)** $2 \times 10^6$ atoms were considered.

However, in particular because the FCC structure corresponds to a multiple cell, it is questionable whether the aspect of the diffraction pattern remains the same when the system involves a limited number of layers. This was simulated for 2 and 4 layers respectively (**FIG. 4b** and **4c**). It is worth noting that 6 spots with the hexagonal symmetry now appear in the first ring, as it was expected for the hexagonal structure only (**FIG. 4d** to **4f**). The intensity of those spots is seen to decrease from the 2-layer to the 4-layer structure, which is consistent with the fact that those spots will not show up at some point as the number of layers increases. Therefore, for diamane and diamanoïd films made of few layers only, it is again not possible to unambiguously discriminate the diffraction patterns of pure $sp^3$–C crystalline phase (diamond or lonsdaleite) from that of few-layer graphene by considering the spot distribution and symmetry.

Finally, we explored the possibility to discriminate between the various structures by measuring spot intensities. Indeed, in [14], the intensity distribution over the diffraction peaks in the electron diffraction patterns of F-diamane and AB-stacked bi-layer graphene was studied and compared to data obtained from simulated diffractograms acquired from DFT-optimized structures. Therefore, we did the same, and analyzed the intensity distribution over the diffraction peaks in the electron diffraction patterns of a single FLG domain after the exposure to the hot-filament-promoted hydrogenation process (**FIG. 5**).

**FIG. 5a** shows typical electron diffraction pattern of a single FLG domain after the exposure to the hot-filament-promoted hydrogenation process. **FIG. 5b-d** display the profile plots of the diffraction peak intensities obtained from pattern regions highlighted by the rectangular boxes in **FIG. 5a**. **FIG. 5b-d** only show three of the six profiles for clarity purpose. For half of the profile lines, the intensity of the spots of the 1$^{st}$ family planes (closest to the center of the pattern) is higher than the intensity of the spots of the 2$^{nd}$ family planes (**FIG. 5b** and **5c**). This is similar to what was reported for F-diamane in [14]. However, for the other half of the intensity profiles, there is no clear trend and the situation differs from the ideal case (**FIG. 5d**).

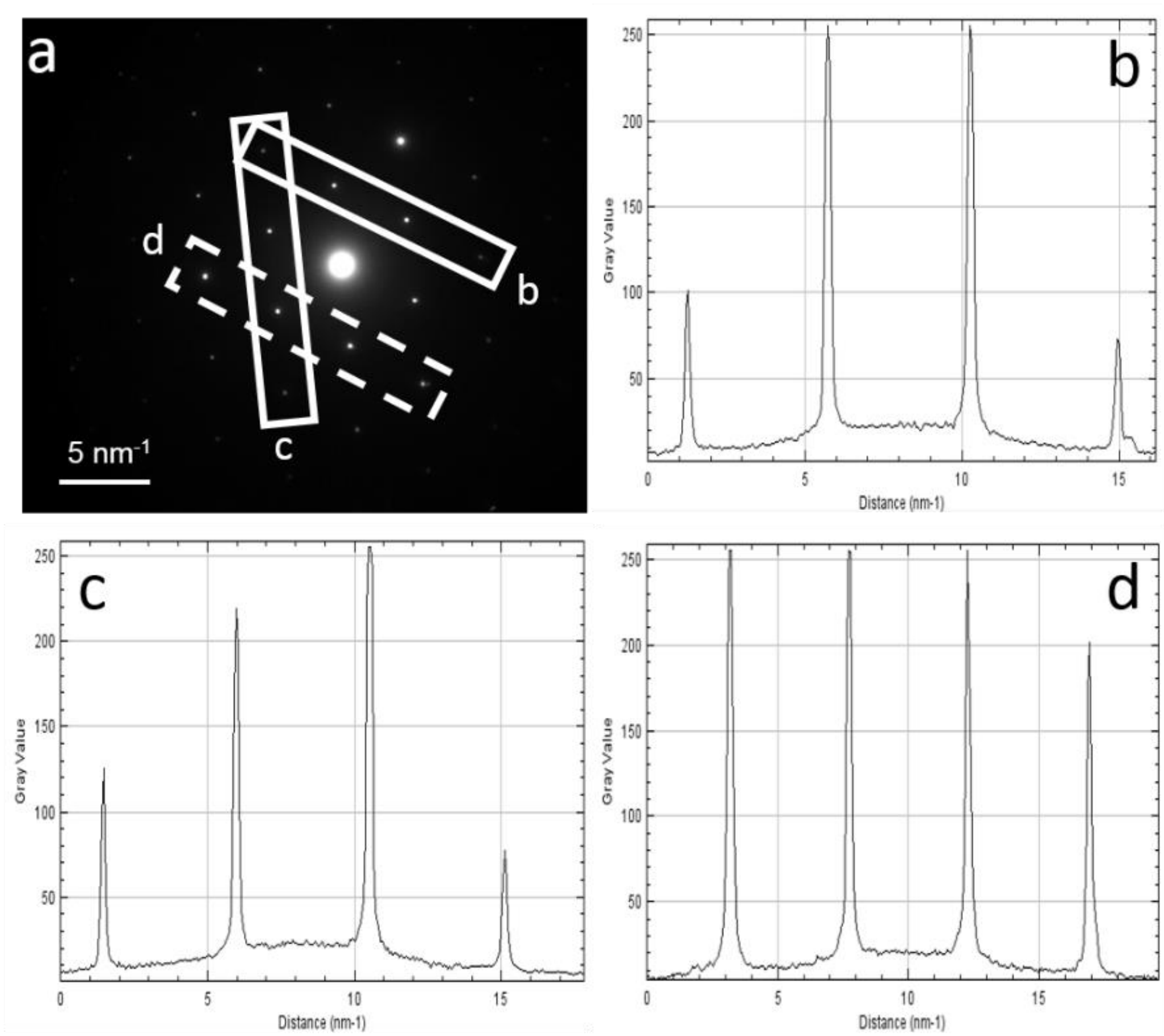

**FIG. 5. (a)** Typical electron diffraction pattern of FLG after the exposure to the hot-filament-promoted hydrogenation process and (**b-d**) profile plots of the diffraction peak intensities obtained from pattern region highlighted by the rectangular boxes in **(a)**. The letters in (**a)** are used to label the profiles shown in (**b-d**).

This shows that such a measurement has a poor significance, because the Ewald sphere intercepts the elongated reciprocal rods at various height with respect to the local film distortion. Indeed, the fact that the coherent domains are ultrathin films makes that they distort easily. Even though the distortions are slight, because of the very small electron wavelength, Bragg angles are in the range of a fraction of degree, hence distortions in a similar range are enough for some family planes be no longer under the Bragg angle. On the other hand, the ultrathin nature of the films makes that the reciprocal nodes are elongated in the z* direction making them reciprocal lines. Therefore, although some scattering events should normally not occur because of the film distortion, the reciprocal line of the family plane involved in the distortion could be elongated enough for touching the Ewald sphere surface. The result is that the scattering occurs anyway, a spot appears in the diffraction pattern which actually is a cross-section of the reciprocal line by the Ewald sphere, but the intensity of the spot cannot be the nominal one because the Ewald sphere does not cross the reciprocal line at its maximum intensity but at various height depending on the distortion. This corresponds to the "interference error" effect well-known by transmission electron microscopists. Therefore, one could not expect discriminate between the various possible structures based on the analysis of spot intensities either.

In some regions of the FLG after exposure to the hot-filament-promoted hydrogenation process, electron diffraction patterns differ from the typical patterns of pristine FLG and/or of pure ultrathin and crystalline $sp^3$–C sheets. **FIG. 6** displays some examples of such patterns taken from different flakes. Over 53 flakes analyzed, such patterns were not observed in pristine FLG. The complex patterns shown in **FIG. 6** including satellite peaks are characteristic of patterns for twisted superimposed coherent domains (TCD) with similar periodicities of small twisted angle. Referring to the model in **FIG. 1**, such TCDs could be the upper diamanoïd domain (represented by L1 and L2, completed with L3) and the lower graphenic domain (represented by L4 and other graphenes underneath) – see the introduction to the model in Section 2.4. Therefore, at the interface between the two twisted domains are two graphene layers (L3 and L4) which are in twisted configuration. Twisted bilayer (TBL) is a system recently studied [26,27] which generates specific Raman signatures [28]. Therefore, **FIG. 6** reveals such an occurrence of TCDs upon hydrogenation.

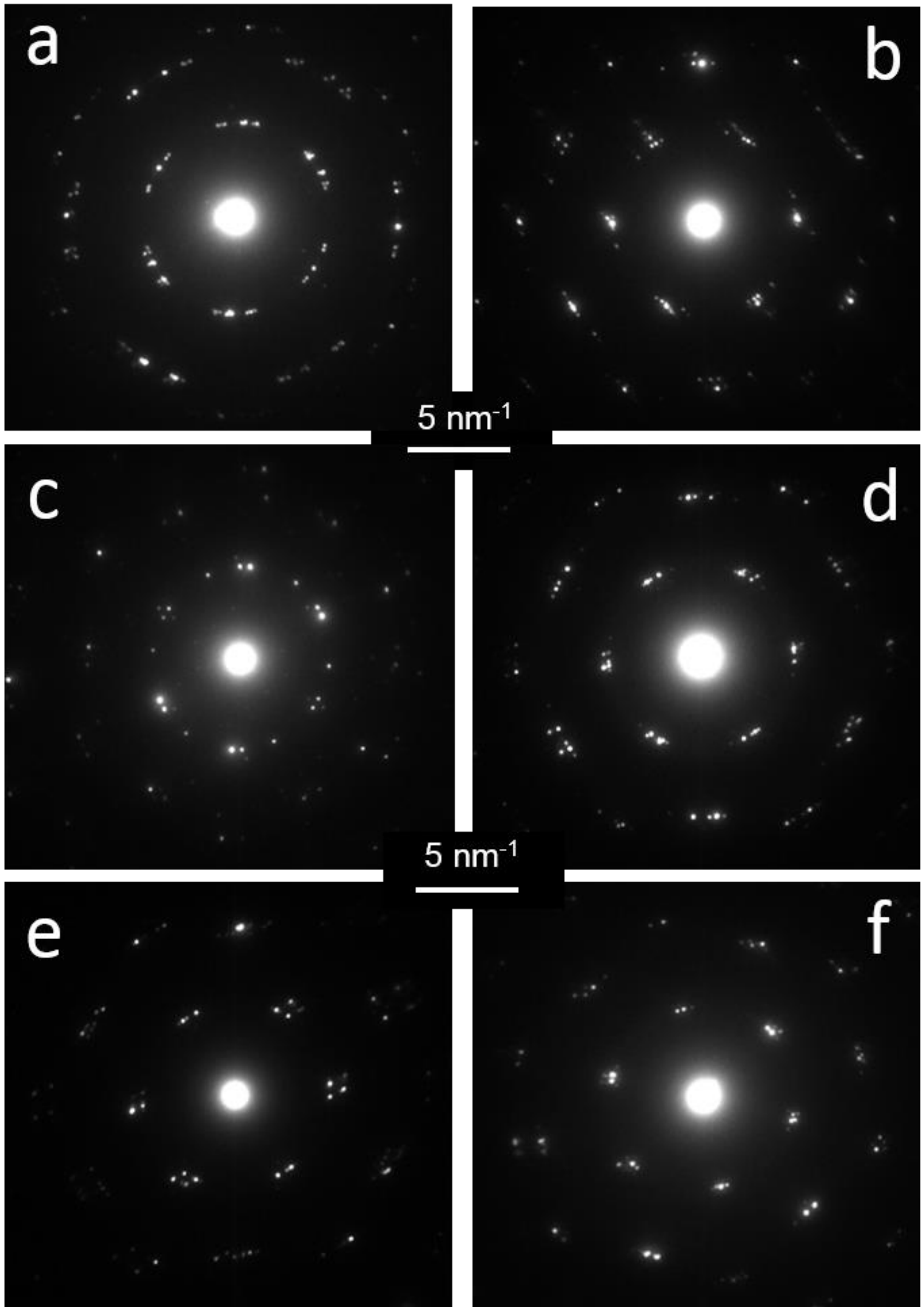

**FIG. 6. (a** to **f)** Typical diffraction patterns of specimen flakes after the exposure to the hot-filament-promoted hydrogenation process.

This interpretation is supported by the observation of moiré patterns in TEM images typical of such twisted systems (**FIG. 7b** and **7d**) and the detection of the specific peaks observed at ~1385 $cm^{-1}$ and in ~1669 $cm^{-1}$ in UV Raman spectra ($TBL_1$ and $TBL_2$ in **FIG. 8**), which were previously reported for TBL [28].

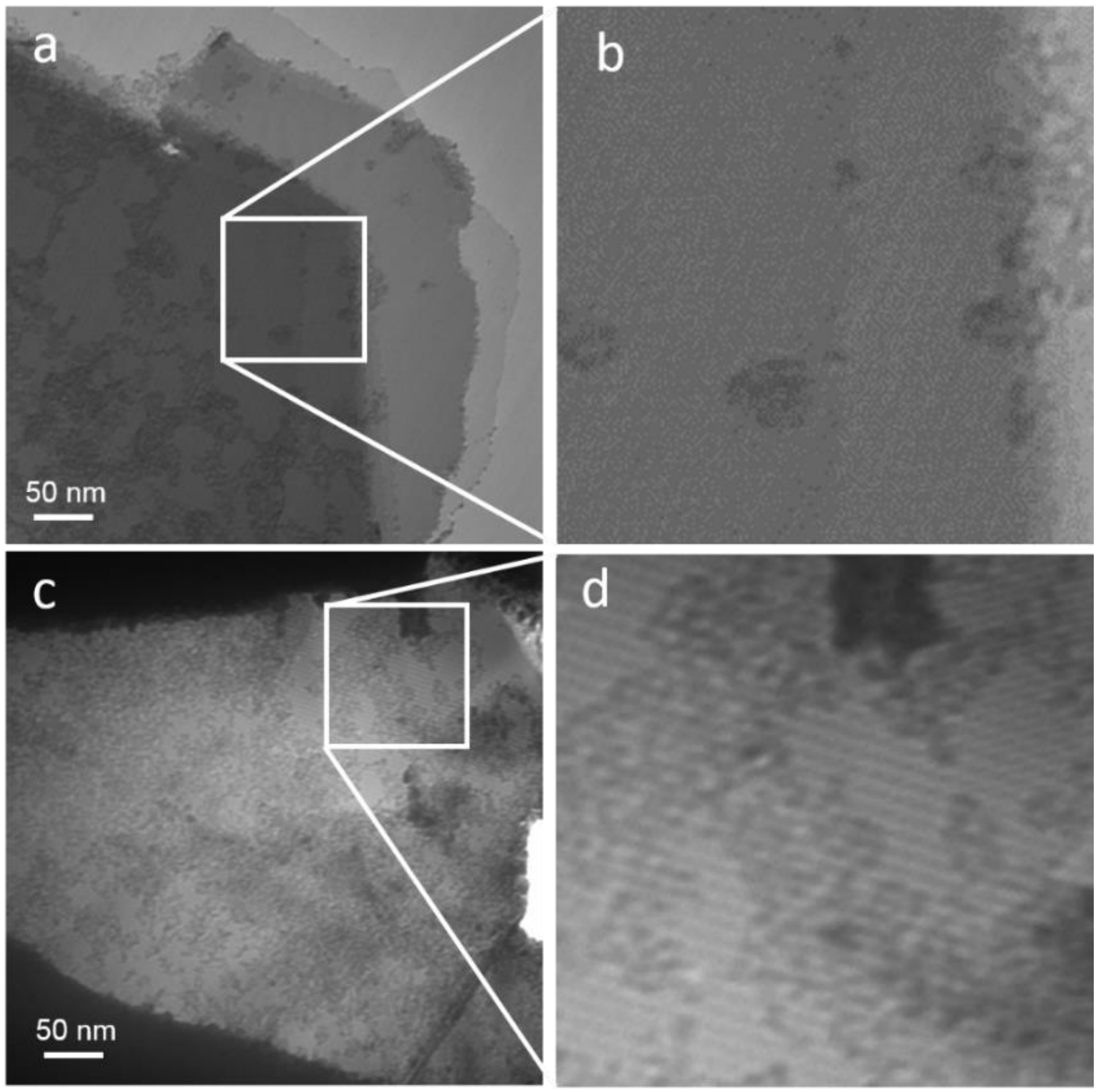

**FIG. 7. (a)** and **(c)** are the TEM images of the specimen areas having provided the diffraction patterns shown in **FIGS. 6e** and **6f**, respectively. **(b)** and **(d)** are magnified views of the squared areas in **(a)** and **(c)**, to better reveal the moiré periodicities.

It is supposed that the slight rotation of the superimposed domains from the TCDs is induced by the relaxation of huge (several GPa) local constraints which develop between superimposed layers as a result of the $sp^2$–C-to-$sp^3$–C conversion, combined with a partial conversion only of the upper layer (thereby consisting of an in-plane hybrid $sp^3$–C–$sp^2$–C layer). In diamond-like carbon, a material that contains a variable fraction of $sp^3$–C and $sp^2$–C, internal stress can reach several GPa [9,10,29].

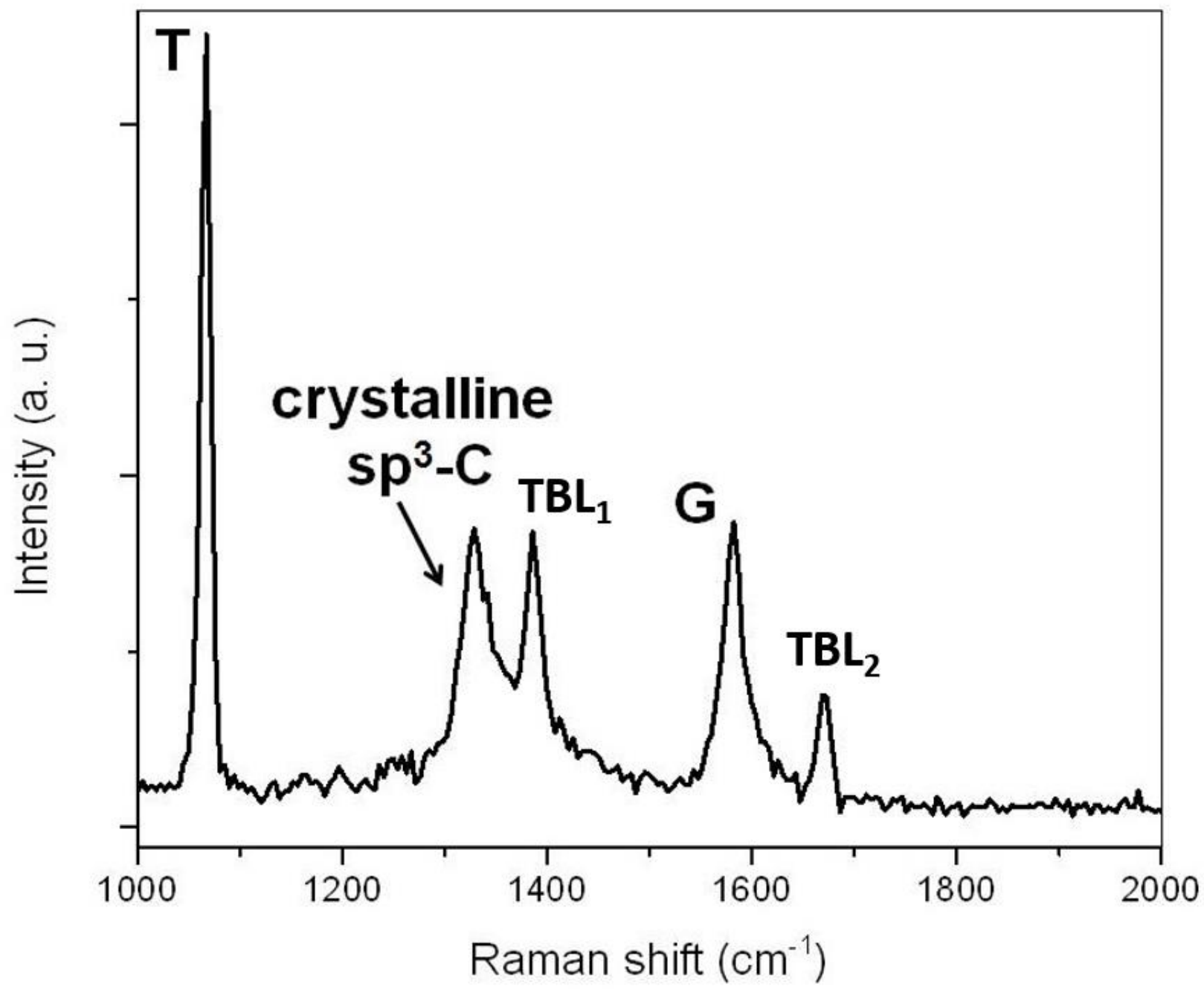

**FIG. 8.** Typical UV Raman spectrum of a specimen flake after the exposure to the hot-filament-promoted hydrogenation process showing peaks typical of twisted bilayer graphene, labeled $TBL_1$ and $TBL_2$. Those peaks are also present in the Raman spectrum in **FIG. 2**. The G band reveals the partial conversion only.

Together with the sharp stretching mode at 1319-1337 $cm^{-1}$ and assigned to lonsdaleite or diamond, UV Raman spectra of hydrogenated samples display a sharp T peak at 1055-1071 $cm^{-1}$ assigned to bonding between $sp^3$–C (**FIG. 2** and **8**) [2]. We further investigated the physical origin of this T peak by using DFT calculations. For pure $sp^2$–C and $sp^3$–C materials, no T peak is present. We have already reported [2] a spatial correlation between the occurrence of the diamond/lonsdaleite stretching mode peak and that of the T peak while there is no spatial correlation between the location of the G peak and that of the T peak. Thus, we have hypothesised that the T peak could be related to the interface between a $sp^2$–C layer and a $sp^3$–C layer, and consequently we have investigated the phonon dispersion of the ABBA-stacking model proposed in **FIG. 1**. Upon full atomic relaxation, all frequencies remain positive, proving the stability of the proposed structure. Three distinct Raman active modes in the range of interest are present, with wavenumbers from 1050 to 1100 $cm^{-1}$ (**FIG. 9**). Normal mode analysis shows that two of them, which correspond to a large rotation of hydrogen atoms (see **SI.2** for atomic vector representation), are degenerated with a wavenumber of 1099 $cm^{-1}$ at the center of the Brillouin zone (Γ point) and are visible in Raman spectra. Interestingly, the mode at 1078 $cm^{-1}$, which involves atoms from L1 and L2, can be associated to a combination of the stretching of the $sp^3$-C bonds with an optical out-of-plane (ZO) mode of graphene membrane. Due to the periodic repetition of the motif in DFT calculations, delamination is not possible and the full cell relaxation leads to compressive strain in the $sp^3$–C part (L1-L2) and to extensive strain in graphene layers (L3-L4, and further layers underneath), explaining the position of the G band which is downshifted by almost 150 $cm^{-1}$ in the calculation compared to the pristine graphene (see **FIG. SI.1**). Using the conversion factor of 4.5 $cm^{-1}$/GPa [30], the stress brought to the graphene domain (L4 and further layers underneath) can be estimated at -33 GPa. Obviously, this is a huge stress value which is by far enough to generate the delamination and twisting events which occur experimentally in the material, although the periodic conditions in the calculation cannot account for them. The system exhibits a perfect, periodic interface contrary to amorphous materials. As a consequence, the experimental linewidth of the T band is as low as 14.0 $cm^{-1}$.

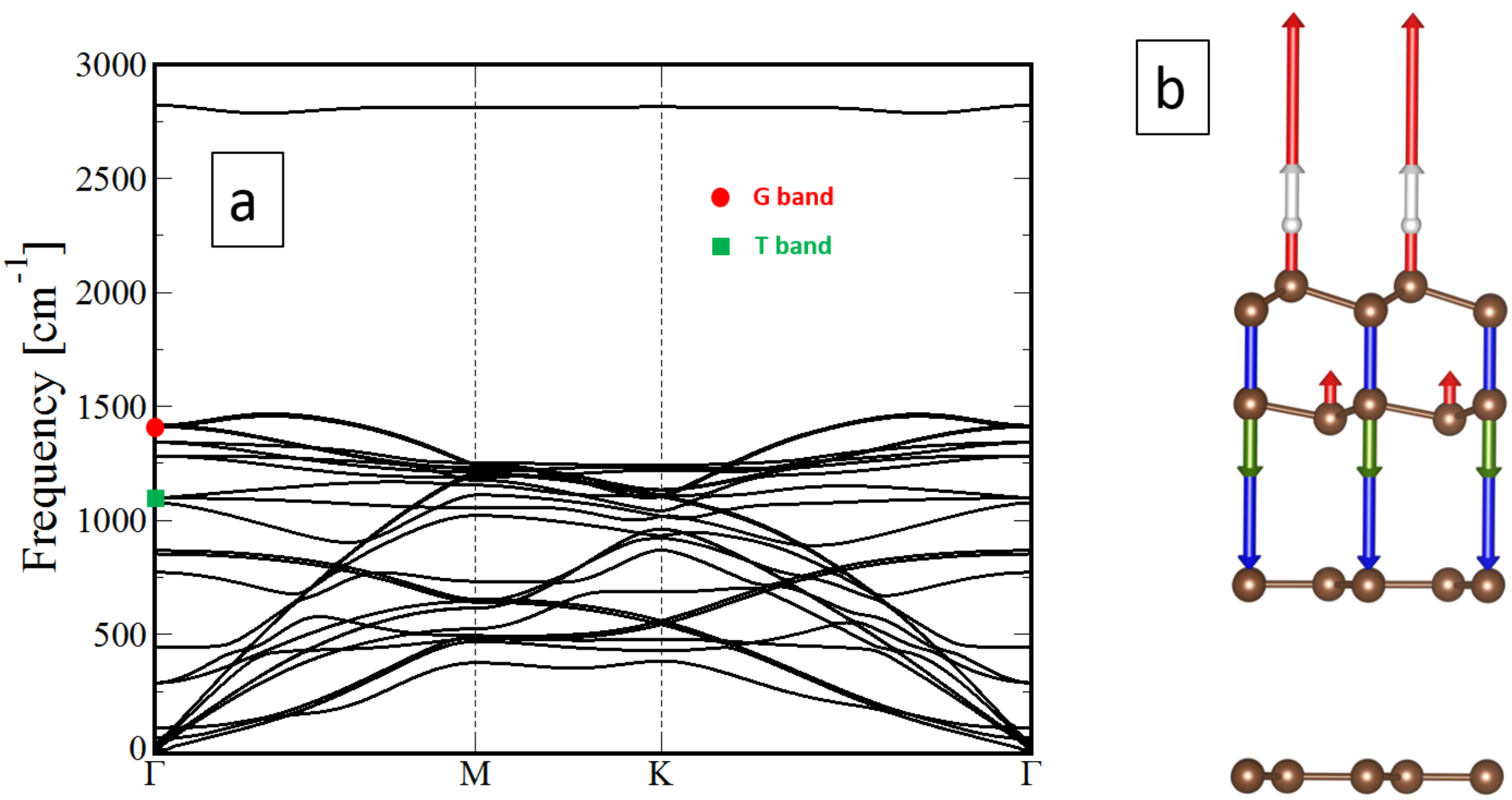

**FIG. 9. (a)** Phonon dispersion curve of the partially hydrogenated ABBA stacked FLG model proposed in **FIG. 1** based on DFPT calculations. **(b)** Corresponding vectors associated to the atomic displacements along the normal mode at 1078 $cm^{-1}$. Vector lengths are not at the scale of the model, but are proportional to each other.

## 4. Conclusion

We reported on new additional evidences, from electron diffraction at 5 keV and Raman spectroscopy, supported by calculations, confirming the successful conversion of few-layer graphene into nanometer-thick and crystalline $sp^3$–C sheets from the exposure of FLG to H radicals produced in a hot-filament process at low temperature and pressure. The results confirmed previous multi-wavelength Raman spectroscopy and FTIR microscopy analysis. Complex electron diffraction patterns differing from pristine FLG and including satellite peaks reveal that twisted coherent domains involving a minimum of one diamanoïd domain and one unconverted graphenic domain underneath are formed, presumably due to the relaxation of stress resulting from the partial $sp^2$–C to $sp^3$–C conversion. A configuration of twisted bilayer graphene (TBL) is created at the interface between the twisted, superimposed domains. Peaks ascribed to TBL are also evidenced in Raman spectra at 1385 $cm^{-1}$ and 1669 $cm^{-1}$, consistently with the recent literature. Also, by using DFT calculations, the exact origin of the T peak was revealed. It originates from a mixt $sp^2$–C-$sp^3$–C interface layer created between the highly hydrogenated $sp^3$–C surface and the unconverted graphene layer(s) underneath which is present. The frequent evidence of $sp^3$–C to $sp^2$–C partial conversion confirms that starting from FLG with a number of layers higher than two is not favourable for producing diamanoïds. It is expected to elaborate large area diamane layer by using Bernal-stacked bi-layer graphene as pristine material. On the other hand, the use of FLG with other stacking order may be used for the fabrication of graphene/diamanoïd heterostructures.

## Acknowledgements

This research was funded by the Ministry of Higher Education, Science and Technology (MESCyT) of the Dominican Republic (2010-2011, 2012, 2015 and 2016-2018 FONDOCyT programs). F.P. and M.M. greatly acknowledge MESCyT and PUCMM for strong financial, administrative and technical support; 2017 NEXT Guest scientist Program and the French Embassy in the Dominican Republic for travel support. I.C.G. acknowledges the Calcul en Midi-Pyrénées initiative CALMIP (Project p0812) for allocations of computer time, as well as GENCI-CINES and GENCI-IDRIS for Grant No. 2019-A006096649.